\documentclass[11pt,letterpaper]{article}
\usepackage[margin=1in]{geometry}
\usepackage[T1]{fontenc}
\usepackage{lmodern,microtype}
\usepackage{amsmath,amssymb,amsthm,mathtools}
\usepackage{enumitem,booktabs,array}
\usepackage{graphicx,xcolor,tikz,float,caption}
\usepackage[section]{placeins}
\usetikzlibrary{arrows.meta,positioning,calc,fit}
\usepackage[numbers,sort&compress]{natbib}
\usepackage{xurl}
\usepackage[colorlinks=true,linkcolor=blue!45!black,citecolor=blue!45!black,
  urlcolor=blue!45!black]{hyperref}
\hypersetup{pdftitle={Optimal Matroid Prophets against Almighty Adversaries and Delegation},
  pdfauthor={Mohammad Hajiaghayi},pdfsubject={Matroid prophet inequalities and delegated choice}}
\setlist{itemsep=2pt,topsep=4pt}
\floatstyle{ruled}
\newfloat{algorithm}{tbp}{loa}
\floatname{algorithm}{Algorithm}
\newtheorem{theorem}{Theorem}[section]
\newtheorem{lemma}[theorem]{Lemma}

\theoremstyle{definition}

\theoremstyle{remark}

\newcommand{\E}{\mathbb E}
\newcommand{\Prb}{\mathbb P}
\newcommand{\R}{\mathbb R}
\newcommand{\Ical}{\mathcal I}
\newcommand{\Bcal}{\mathcal B}
\newcommand{\Acal}{\mathcal A}
\newcommand{\OPT}{\operatorname{OPT}}
\newcommand{\ALG}{\operatorname{ALG}}
\newcommand{\exante}{\operatorname{EA}}
\newcommand{\cl}{\operatorname{cl}}
\newcommand{\SFM}{\operatorname{SFM}}
\newcommand{\BR}{\operatorname{BR}}
\newcommand{\ind}{\mathbf 1}

\title{\vspace{-1.0em}Optimal Matroid Prophets against Almighty Adversaries\\
  and Delegation}
\author{Hamed Abdi\thanks{University of Maryland, College Park. Email: \texttt{hamedabdi37@gmail.com}.}
\and Kiarash Banihashem\thanks{University of Maryland, College Park. Email: \texttt{kiarash@umd.edu}.}
\and MohammadTaghi Hajiaghayi\thanks{University of Maryland, College Park. Email: \texttt{hajiaghayi@gmail.com}.}
\and Danny Mittal\thanks{University of Maryland, College Park. Email: \texttt{dannymittal@gmail.com}.}}

\begin{document}
\maketitle

\begin{abstract}
We prove an optimal one-half matroid prophet inequality against an
almighty ordering adversary: the adversary observes every realized
value before choosing the arrival order, whereas the algorithm sees
values only as they arrive and must accept irrevocably. The algorithm
is deterministic, and its guarantee holds against the stronger ex-ante
relaxation. Before any values are revealed, it partitions the ground set,
replaces the matroid by a stricter direct sum of minors, and fixes one
threshold for each part.
It then accepts greedily subject to these fixed tests.

The main lemma bounds the expected value of the least valuable basis
of the eligible elements. An element contributes either through the
rank spanned by eligible elements or through excess value that every
eligible basis must retain. This bound holds simultaneously for all
bases and therefore permits the arrival order to depend on the entire
realization.

The same construction gives a deterministic, agent-oblivious mechanism
for (possibly correlated) matroid delegation, improving the $1/4$ guarantee
to the optimal $1/2$ factor. The principal values are independent,
while the strictly positive agent utilities may have arbitrary dependence
on the principal values and on one another. The guarantee holds for every best
response, including ties resolved against the principal. Both factors
are tight already in rank one. For rational finite-support distributions,
preprocessing takes polynomial time; each eligible online arrival
requires at most one query to the original matroid's independence oracle.
\end{abstract}

\section{Introduction}

Prophet inequalities ask how much is lost when random values must be
selected before all of them are known. In the matroid version, each
element has an independent nonnegative value drawn from a known
distribution. Values are revealed one at a time, and the algorithm
must maintain an independent set through irrevocable decisions. Its
benchmark is the expected maximum-weight independent set after all
values have been revealed. The classical one-half guarantee extends
from selecting one element to arbitrary matroids
\cite{samuelcahn1984,kleinbergweinberg2012}.

The order of revelation matters as much as the feasibility constraint.
An adversary that chooses the next element using only previously
revealed values leaves unseen values independent of the history.
An \emph{almighty adversary} has no such restriction: it sees the full
value vector and then chooses the worst arrival order. An element's
position can therefore reveal information about its own value and the
values of elements that have not yet arrived. In particular, one cannot
assume that an element's value is independent of its availability when
it is presented.

This stronger model also arises naturally in delegation. A principal
commits to a menu of acceptable solutions, and an informed agent
chooses a solution according to its own utilities. If principal and
agent utilities are correlated, presenting elements in decreasing
agent utility produces an order that depends on the principal's
realized values. Greedy prophet rules against an almighty adversary
therefore give delegation mechanisms
\cite{bechtel2021,banihashem2025}. For a general matroid, the greedy
contention-resolution approach gives a one-quarter guarantee
\cite{feldman2021,bechtel2021,banihashem2025}, whereas the two-element
rank-one obstruction gives an upper bound of one half
\cite[Proposition~5.3]{bechtel2021}.

We obtain the sharp one-half guarantee in both settings. The proof
does not analyze the order chosen by an adversary or the preferences
used by an agent. Instead, preprocessing constructs a family of
acceptable sets for which \emph{every inclusion-wise maximal eligible
set} has sufficient expected value, even when the least valuable such
set is chosen separately in every realization. This common statement
is the reason the two applications admit the same sharp constant.

\subsection{Our results}

Let $M=(E,\Ical)$ be a finite matroid. Each item $i\in E$ has an
independent value $X_i\geq0$ with known law $D_i$. For a set $S$, write
$X(S)=\sum_{i\in S}X_i$, and let
$\OPT_M(X)=\max_{I\in\Ical}X(I)$. We assume
$\E[\OPT_M(X)]<\infty$. Let $\exante(M,D)$ denote the standard
ex-ante relaxation, defined in Section~\ref{sec:preliminaries}; it
upper-bounds $\E[\OPT_M(X)]$.

\begin{theorem}[Optimal almighty matroid prophet inequality]
\label{thm:prophet}
There is a deterministic online algorithm, depending only on $M$ and
the marginal laws $D_i$, such that
\begin{equation}
 \E_X\!\left[\min_{\pi\in\mathfrak S_E}\ALG(X,\pi)\right]
 \geq \frac12\exante(M,D)
 \geq \frac12\E[\OPT_M(X)].
 \label{eq:main-prophet}
\end{equation}
\end{theorem}

The algorithm fixes a direct sum $\widehat M$ of matroid minors and
one threshold per summand. Every independent set of $\widehat M$ is
independent in $M$. It accepts an arriving nonloop item if it meets
its threshold and preserves independence in $\widehat M$; it rejects
original loops.

The minimum in~\eqref{eq:main-prophet} is \emph{inside} the expectation.
Thus the adversary may choose a different permutation for every
realization. The algorithm has no private random seed. The theorem
allows atoms, unequal distributions, and unbounded supports; finite
expected optimum is the only integrability assumption.

In delegation, the principal and the agent have additive utilities
$X(S)$ and $Y(S)=\sum_{i\in S}Y_i$. The informed agent proposes a set
after observing all utilities. The principal commits to a menu in
advance, verifies only submitted identities and principal values,
and implements accepted proposals in full. Rejection or no agreement
yields the empty set. A menu is \emph{agent-oblivious} if it uses
neither distributions nor realizations of agent utilities. Write
$\BR(X,Y)$ for the set of implemented best-response outcomes.

\begin{theorem}[Optimal deterministic correlated delegation]\label{thm:delegation}
Suppose the principal values $(X_i)_{i\in E}$ are mutually
independent, nonnegative, and almost surely finite, with known
marginal laws $D_i$ and
$\mathbb E[\operatorname{OPT}_M(X)]<\infty$.
The agent utilities satisfy $0<Y_i<\infty$ almost surely
and may depend arbitrarily on the principal values and on
one another.

There is a deterministic, agent-oblivious, proposal-only menu
satisfying
\begin{equation}
\mathbb E\!\left[
  \min_{S\in\operatorname{BR}(X,Y)} X(S)
\right]
\ge \frac12\,\operatorname{EA}(M,D)
\ge \frac12\,\mathbb E[\operatorname{OPT}_M(X)].
\label{eq:main-delegation}
\end{equation}
\end{theorem}

The menu accepts a proposal exactly when its items meet their
thresholds and its set is independent in $\widehat M$. The vector $Y$
may depend arbitrarily on $X$ and across coordinates.
In particular, the result includes independent item pairs $(X_i,Y_i)$
with arbitrary within-item correlation. The stronger formulation in
Theorem~\ref{thm:delegation} costs nothing: the proof uses independence
only among principal values. Strict positivity of the agent utilities
ensures that every best response is maximal, even with adversarial
tie breaking. No assumption on their expectations is needed.

\paragraph{Optimality and computation.}
The factor one half is tight already for single-choice delegation
\cite[Appendix B, Proposition 1]{kleinbergkleinberg2018};
see also \cite[Proposition 5.3]{bechtel2021}
for a two-item rank-one construction.
The factor in Theorem~\ref{thm:delegation} matches the rank-one
lower bound of Bechtel and Dughmi~\cite[Proposition~5.3]{bechtel2021}
for deterministic proposal-only menus, even when the principal knows
and can verify agent utilities.\footnote{Their two-item construction
assigns agent utility zero to both outcomes of the risky item and
one to the safe item. Replacing the former utilities by any fixed
$\eta\in(0,1)$ preserves the deterministic lower bound and makes
all agent utilities strictly positive.}
Both bounds compare with the expected offline optimum and hence also
apply to the ex-ante benchmark. Thus correlated matroid delegation
attains the rank-one bound, closing the $1/4$ versus $1/2$
gap for this deterministic model.

For explicitly listed rational finite-support distributions and an
independence oracle for $M$, preprocessing takes polynomial time.
If $n$ is the number of nonloop items and $L$ is the input description
length, it makes $O(n(L+\log\max\{1,n\}))$ submodular minimizations. Each eligible
arrival subsequently uses at most one original independence query.
The delegation verifier uses at most one such query for each nonempty
submitted block. Section~\ref{sec:implementation} gives exact
pseudocode, including threshold recovery and deterministic treatment
of all ties.

\subsection{Related work and the role of the model}

\paragraph{Prophets and contention resolution.}
Kleinberg and Weinberg~\cite{kleinbergweinberg2012} obtain the sharp
one-half matroid prophet inequality with thresholds that depend on
the accepted set. Lee and Singla~\cite{leesingla2018} establish the
corresponding ex-ante guarantee for arrival orders fixed before the
values are realized, and optimal online contention resolution when
that fixed order is known. Feldman, Svensson, and
Zenklusen~\cite{feldman2021} develop greedy online contention resolution,
which tolerates an almighty adversary and yields the one-quarter
matroid prophet guarantee. The distinction between weighted prophet
guarantees and per-element contention resolution is essential here.
Our theorem controls expected total value; it does not promise a
uniform conditional selection probability for every element.
In particular, it does not give a one-half greedy online contention
resolution scheme: Livanos~\cite{livanos2022} proves a $1/e$ upper bound
for that stronger per-element requirement already in rank one.
Dinev and Weinberg~\cite{dinev2024} further study the role of the
adversary for uniform-matroid contention resolution.

\paragraph{Fixed thresholds and feasibility.}
Our thresholds are fixed, but feasibility is tested in a predetermined
matroid $\widehat M$ whose independent sets form a subfamily of those
of $M$. This distinction separates the algorithm from rules that
use fixed item thresholds and always accept when feasible in the
original matroid. The latter restriction is studied by Chawla,
Goldner, Karlin, and Miller~\cite{chawla2024} and
Jiang~\cite{jiang2026}. Restricting the acceptable family is part of
the greedy contention-resolution framework~\cite{feldman2021}.
Almost non-adaptive matroid prophet algorithms are also studied by
Kalantarzadeh and Pashkovich~\cite{kalantarzadeh2026}. The guarantee
proved here is explicitly for the realization-dependent worst order
in~\eqref{eq:main-prophet}; the fixed-threshold form alone does not
establish a guarantee against this adversary.
Figure~\ref{fig:reservation} gives a concrete example of why the
additional constraints matter for our result. The comparator always
remains the optimum in $M$.

\paragraph{Delegation.}
Armstrong and Vickers~\cite{armstrongvickers2010} study delegated
project choice, and Kleinberg and Kleinberg~\cite{kleinbergkleinberg2018}
relate approximate delegation to threshold stopping rules.
Bechtel and Dughmi~\cite{bechtel2021} extend the connection to delegated
stochastic probing and show how greedy almighty prophet algorithms
give deterministic delegation policies. With a matroid selection
constraint and no probing constraint, their guarantee is $1/4$.
Banihashem, Hajiaghayi, Krysta, and Shin~\cite{banihashem2025} study
delegated choice under combinatorial constraints, including the
correlated-utility model and its matroid connection. We use this
established connection; the direct proof in
Section~\ref{sec:delegation} makes clear why all best responses and
arbitrary utility correlations are covered. The contribution is the
sharp matroid guarantee, rather than a new general reduction between
the two problems. We impose no outer probing constraint, sampling
cost, or transfer.

\paragraph{Other information and feasibility models.}
MacRury, Nuti, and Vondr\'ak~\cite{macrury2026} study ex-ante prophet
inequalities and contention resolution for matchoids under fixed
adversarial and uniformly random arrival orders.
In the single-sample model, Abdi et al.~\cite{abdietal2026} obtain an
optimal one-half matroid prophet guarantee for orders fixed
independently of samples, values, and the algorithm's randomness.
An OpenAI preprint~\cite{openai2026} gives a constant one-sample
guarantee when an almighty adversary sees samples, values, and the
full random seed. Our result assumes known distributions and achieves
the exact one-half factor in polynomial time for explicit finite
supports. The proof is specific to matroid structure: it uses equal
cardinality of maximal independent sets, closure, and contraction.
It does not assert the same guarantee for arbitrary downward-closed
constraints or a new secretary guarantee.

\paragraph{Organization.}
Section~\ref{sec:techniques} explains the proof and the delegation
mechanism. Section~\ref{sec:preliminaries} fixes notation and the
benchmark. Sections~\ref{sec:decomposition} and~\ref{sec:basis}
prove the prophet theorem, and Section~\ref{sec:delegation}
proves the delegation guarantee.
Section~\ref{sec:implementation} supplies the polynomial-time
implementation.

\section{Our approach and techniques}
\label{sec:techniques}

\subsection{Protecting every eligible basis}

\paragraph{Decompose first, then run blockwise greedy.}
Different parts of the matroid may have different value scales.
After discarding loops, we partition $E$ into disjoint \emph{blocks}
$U_1,\ldots,U_m$. Write $F_0=\varnothing$ and
$F_j=U_1\cup\cdots\cup U_j$ for their prefixes. On each block we use
the matroid $N_j=(M/F_{j-1})|U_j$: contract the earlier blocks, then
restrict to $U_j$. This reserves rank for earlier blocks. We choose
these blocks and thresholds $t_1\geq\cdots\geq t_m\geq0$ using only
the value distributions, as sketched below.

The algorithm runs greedy separately in each $N_j$: an arriving item
$i\in U_j$ is \emph{eligible} if $X_i\geq t_j$, and is accepted if
adding it preserves independence in $N_j$. Equivalently, it maintains
independence in the direct sum $\widehat M=\bigoplus_jN_j$, whose
independent sets are exactly those $S$ with
$S\cap U_j$ independent in $N_j$ for every $j$. Every such set is
also independent in $M$.

\paragraph{Protect every possible blockwise outcome.}
Blockwise greedy ends with a basis of the eligible set in each $N_j$.
An almighty adversary can use the realized values to influence which
bases are selected, allowing low-valued eligible items to displace
more valuable ones. We therefore bound the expected value of the
\emph{least valuable} eligible basis in each block. This local bound
applies to every arrival order; the decomposition will ensure that
the bounds across blocks add up to half the original benchmark.

\paragraph{Balance threshold value and expected excess.}
For item $i$, define its expected excess above $t$ by
\[
 e_i(t)=\E[(X_i-t)_+].
\]
Think of an accepted value as a threshold payment plus excess. In a
rank-one instance, the natural balance is
$\sum_i e_i(t)=t$: the threshold and the total expected excess have
the same size. The elementary inequality
$X_i\leq t+(X_i-t)_+$ then bounds the expected maximum by $2t$.
To achieve one half, it remains to prove that even the least valuable
maximal eligible set has expected value at least $t$, taking its value
to be zero when no item is eligible.

To analyze a single block, write $N=N_j$, $U=U_j$, and $t=t_j$.
The corresponding balance conditions are
\begin{equation}
 e(A,t)\leq t r_N(A)\quad(A\subseteq U),
 \qquad e(U,t)=t r_N(U),
 \label{eq:overview-balance}
\end{equation}
where $e(A,t)=\sum_{i\in A}e_i(t)$. The subset inequalities say that
no part of the block carries more expected excess than its rank can
pay for. Tightness on the entire block gives the desired balance.
We will show that these conditions guarantee expected value at least
$e(U,t)=t r_N(U)$ even for the least valuable basis of eligible items.

\paragraph{What every basis must retain.}
Let $R=\{i\in U:X_i\geq t\}$ be the eligible set and let
$F=\cl_N(R)$ be its closure. Every basis of $R$ has $r_N(F)$ elements,
each worth at least $t$. This gives a common contribution
$t r_N(F)$, regardless of which basis is selected.

\begin{figure}[H]
\centering
\begin{tikzpicture}[x=1.05cm,y=1.05cm,
  vertex/.style={circle,fill=black,inner sep=2.1pt},
  lab/.style={font=\small,align=center}]
  \coordinate (a) at (0,1.9);
  \coordinate (b) at (0,0);
  \coordinate (c) at (2.1,0.95);
  \coordinate (d) at (4.2,0.95);
  \coordinate (e) at (6.3,0.95);
  \draw[thick] (a)--(b)--(c)--(a);
  \draw[very thick,blue!55!black] (c)--(d);
  \draw[dashed,thick] (b) to[bend right=23] (d);
  \draw[densely dotted,thick] (d)--(e);
  \foreach \v in {a,b,c,d,e}{\node[vertex] at (\v) {};}
  \node[lab,above=3pt] at (3.15,0.95) {forced item $i$};
  \node[lab] at (1.05,2.45) {eligible cycle};
  \node[lab] at (2.1,-0.7) {ineligible, but spanned};
  \node[lab] at (5.35,0.05) {outside the closure};
  \node[lab,anchor=west,text width=3.6cm] at (7.0,1.0)
    {Every eligible basis\\keeps the bridge.\\[5pt]
     The cycle edges\\can replace one another.};
\end{tikzpicture}
\caption{Forced elements in a graphic matroid: every basis of the
eligible edges must contain each bridge. Solid edges are eligible.
The dashed edge is ineligible but belongs to their closure because its
endpoints are already connected. The dotted edge joins the isolated
rightmost vertex and lies outside that closure. The solid bridge
belongs to every basis of the eligible set.}
\label{fig:forced}
\end{figure}
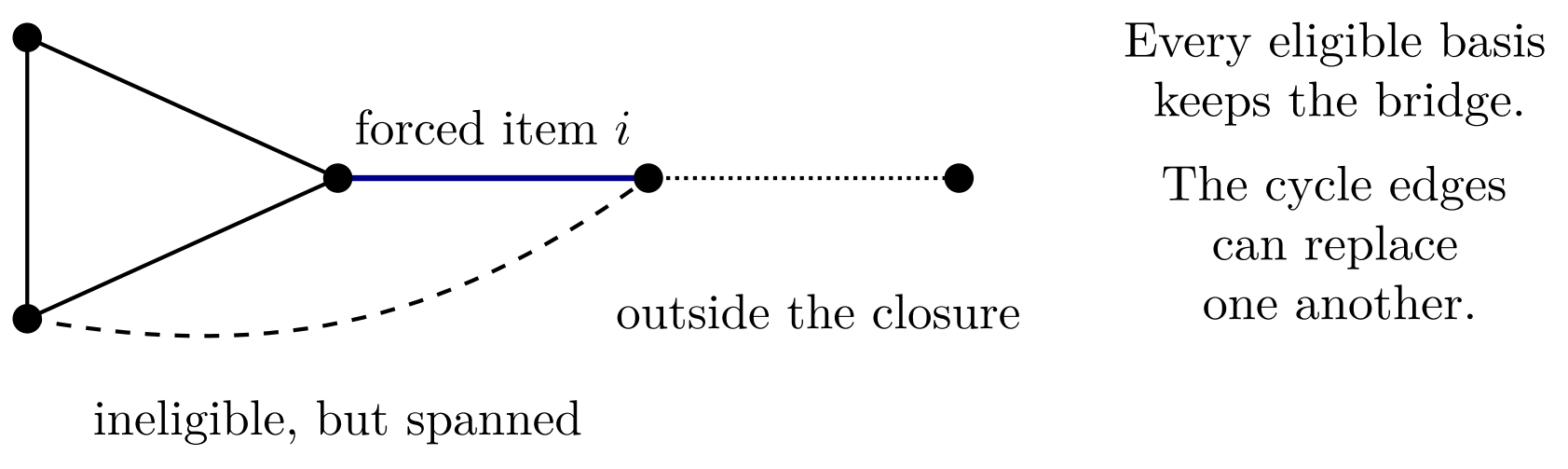

Some eligible elements also contribute excess that cannot be lost.
For each $i$, consider the event
\[
 H_i=\{i\notin\cl_N(R\setminus\{i\})\}.
\]
If $X_i>t$ and $H_i$ occurs, the other eligible elements do not span
$i$. Every basis must then contain $i$, including a basis selected
after seeing all values. Thus, simultaneously for every basis $B$ of
$R$,
\begin{equation}
 X(B)\geq t r_N(F)+\sum_{i\in U}(X_i-t)_+\ind_{H_i}.
 \label{eq:overview-pathwise}
\end{equation}
In a graphic matroid, these forced elements are bridges in the graph
formed by the eligible edges; see Figure~\ref{fig:forced}.

The event $H_i$ was defined after removing $i$. It therefore depends
only on other values and is independent of $X_i$. This is the only
independence needed in the local argument. Taking expectations
in~\eqref{eq:overview-pathwise} replaces the excess term by
$\sum_i e_i(t)\Prb(H_i)$.

Why is this enough? If $i$ is outside the full closure $F$, it is
certainly outside the closure after deleting itself. Hence
$\Prb(H_i)\geq\Prb(i\notin F)$. The expected excess credited through
the events $H_i$ is therefore at least
$e(U,t)-\E[e(F,t)]$. The part that may be missing lies inside $F$,
where~\eqref{eq:overview-balance} gives
$e(F,t)\leq t r_N(F)$. This is exactly what the threshold contribution
already pays. The two terms cancel, yielding the promised
\emph{single-block basis bound}:
\begin{equation}
 \E\!\left[\min_{B\text{ basis of }R}X(B)\right]\geq e(U,t).
 \label{eq:overview-basis-bound}
\end{equation}
Neither the chosen basis nor the random closure has been assumed
independent of any item value. In particular, the argument survives
an adversary's complete knowledge of the realization.

\paragraph{Find a block at the right value scale.}
We now sketch how to choose the blocks and thresholds above.
At step $j$, the residual matroid is $N=M/F_{j-1}$, starting with
$N=M$. For a nonempty set $A$ in this loopless matroid, let
$\tau_N(A)$ solve
$e(A,t)=t r_N(A)$. Choose the largest of these roots, denoted by $t$.
At that threshold every set satisfies the subset inequality in
\eqref{eq:overview-balance}, and at least one nonempty set is
\emph{tight}, meaning $e(A,t)=t r_N(A)$.
The function
\[
 A\longmapsto t r_N(A)-e(A,t)
\]
is nonnegative and submodular. Its zero sets are closed under union,
so there is a unique largest tight set $U$. Closing a tight set cannot
increase its rank or decrease its excess; the subset inequality
therefore forces its closure to be tight too. Thus $U$ is a flat.

Set $U_j=U$, $t_j=t$, and $N_j=N|U_j$, update
$F_j=F_{j-1}\cup U_j$, and continue in $N/U_j$. Contracting $U_j$
reserves its rank before later, lower-threshold blocks are considered.
The contracted matroid remains loopless. Subtracting the equality
for $U$ from the inequality for $U\cup A$ shows that the next
threshold is no larger. Repetition produces a chain of flats
$F_1\subset\cdots\subset F_m=E$ and decreasing thresholds
$t_1\geq\cdots\geq t_m$.

The resulting direct sum can reject a set that is independent in the
original matroid. This restriction is deliberate: it prevents
lower-threshold items from consuming rank reserved for a block with
larger expected excess. Conversely, every set allowed by the direct
sum is independent in the original matroid. Feasibility follows
directly from the definition of contraction, even if an earlier
block does not fill its reserved rank in a particular realization.

\paragraph{Two equal budgets give one half.}
Let $d_j=r_M(F_j)-r_M(F_{j-1})$ be the rank reserved for block $j$ and
put $K=\sum_j t_jd_j$. Tightness of the blocks says that the total
expected excess is also $K$. Summing the single-block basis
bound~\eqref{eq:overview-basis-bound} gives expected reward at least
$K$, simultaneously over all choices of eligible bases.

To upper-bound the original benchmark, give every item in block $j$
price $t_j$. Any independent set in the original matroid has total
price at most $K$. Indeed, its intersection with every prefix flat
$F_j$ has size at most $r_M(F_j)$; summation by parts and decreasing
prices give the claim. The same calculation holds for every point
in the matroid polytope, so it bounds the ex-ante relaxation as well.
An ex-ante solution pays at most $K$ in prices and at most $K$ in
expected excess. Thus its value is at most $2K$, while even the least
valuable eligible set that is maximal in $\widehat M$ has expected
value at least $K$.

\subsection{From the basis guarantee to delegation}

The delegation mechanism uses exactly the same precomputed blocks
and thresholds. A proposal is accepted if every submitted value
meets its threshold and the submitted set is independent in the
direct sum. These tests involve only the proposal. In particular,
the principal does not need to know which omitted elements were
eligible, what their values were, or how the agent ranks them.

Fix any realization and consider a best response. If the proposal
is not maximal among eligible independent sets, matroid augmentation
allows an eligible item to be added. The extra item has strictly
positive agent utility, so the enlarged proposal is strictly better
for the agent. This contradicts optimality. Every best response
therefore contains a basis of the eligible set in each block.

The bound applies to every eligible basis, so it also covers the
agent's choices. The agent may
select a basis using every principal value, every agent value, and
any tie rule. All of these choices were included in the minimum
in~\eqref{eq:overview-basis-bound}. No independence assumption on
agent utilities is used. The principal's expected utility is at
least $K$, while the ex-ante optimum is at most $2K$.

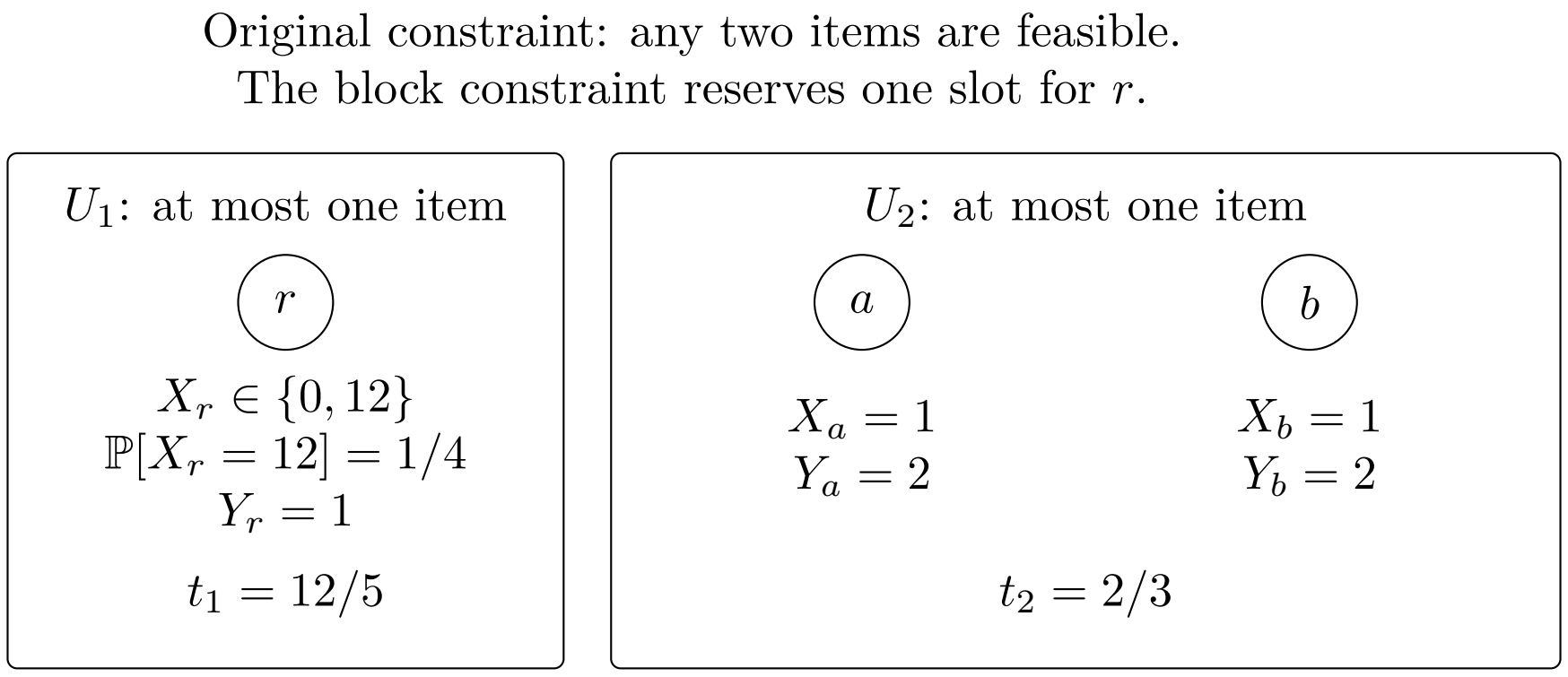
\begin{figure}[H]
\centering
\begin{tikzpicture}[font=\small,
  box/.style={draw,rounded corners=2pt,align=center,inner sep=9pt},
  item/.style={draw,circle,minimum size=7mm,inner sep=0pt}]
  \node[box,minimum width=4.1cm,minimum height=3.8cm] (left) at (0,0) {};
  \node[box,minimum width=7.0cm,minimum height=3.8cm] (right) at (5.9,0) {};
  \node at (0,1.5) {$U_1$: at most one item};
  \node at (5.9,1.5) {$U_2$: at most one item};
  \node[item] at (0,0.8) {$r$};
  \node[item] at (4.25,0.8) {$a$};
  \node[item] at (7.55,0.8) {$b$};
  \node[align=center] at (0,-0.32) {$X_r\in\{0,12\}$\\
    $\Prb[X_r=12]=1/4$\\$Y_r=1$};
  \node[align=center] at (4.25,-0.27) {$X_a=1$\\$Y_a=2$};
  \node[align=center] at (7.55,-0.27) {$X_b=1$\\$Y_b=2$};
  \node at (0,-1.35) {$t_1=12/5$};
  \node at (5.9,-1.35) {$t_2=2/3$};
  \node[align=center] at (3.0,2.6)
    {Original constraint: any two items are feasible.\\
     The block constraint reserves one slot for $r$.};
\end{tikzpicture}
\caption{Why the additional block constraints matter. The original
matroid is $U_{2,3}$. With only the original rank-two constraint, the
agent chooses both safe items and the principal receives $2$.
The displayed menu permits at most one safe item, so every best
response also takes $r$ whenever $X_r=12$. The expected principal
value becomes $4$, compared with an offline optimum of $19/4$.}
\label{fig:reservation}
\end{figure}

Figure~\ref{fig:reservation} illustrates the economic role of
contraction. The risky item has large principal upside but smaller
agent utility than either safe item. The original rank-two constraint
lets the two safe items displace the risky item even when its value
is large. Reserving one rank unit for the risky block rules out that
choice. The verifier does not force the agent to report every
eligible item; strict positivity supplies maximality through the
agent's incentives.

\section{Model and preliminaries}
\label{sec:preliminaries}

\subsection{Matroid notation}

A matroid is a pair $M=(E,\Ical)$, where $E$ is a finite set and
$\Ical\subseteq 2^E$ satisfies the following axioms: $\varnothing\in\Ical$;
every subset of a member of $\Ical$ belongs to $\Ical$; and if
$I,J\in\Ical$ with $|I|<|J|$, then $I\cup\{j\}\in\Ical$ for some
$j\in J\setminus I$. Members of $\Ical$ are called independent sets.
For $A\subseteq E$, define the rank function and closure operation as
\[
 r_M(A)=\max\{|I|:I\subseteq A,\ I\in\Ical\},
 \qquad
 \cl_M(A)=\{i\in E:r_M(A\cup\{i\})=r_M(A)\}.
\]
It is well-known that the rank function is monotone and submodular.
A basis of $A$ is an inclusion-wise maximal independent subset of $A$.
The augmentation axiom implies that every basis of $A$ has size $r_M(A)$
and that every independent subset of $A$ extends to a basis of $A$.
Write $\Bcal_M(A)$ for the family of bases of $A$; in particular,
$\Bcal_M(\varnothing)=\{\varnothing\}$.
A loop is an element $i$ with $r_M(\{i\})=0$.
A flat is a set $F$ satisfying $\cl_M(F)=F$.

For $U\subseteq E$, the restriction $M|U$ is the matroid on $U$ whose
independent sets are the members of $\Ical$ contained in $U$.
For $F\subseteq E$, the contraction $M/F$ is the matroid on
$E\setminus F$ with rank function
\[
 r_{M/F}(A)=r_M(F\cup A)-r_M(F)
 \qquad(A\subseteq E\setminus F).
\]
Equivalently, if $C$ is a basis of $F$, a set $A\subseteq E\setminus F$
is independent in $M/F$ if and only if $C\cup A$ is independent in $M$.
The rank formula gives
\[
 \cl_{M/F}(A)=\cl_M(F\cup A)\setminus F.
\]
In particular, contracting a flat leaves a loopless matroid, and if
$U$ is a flat of $M/F$, then $F\cup U$ is a flat of $M$.
For matroids $N_j$ on disjoint sets $U_j$, their direct sum
$\bigoplus_jN_j$ has ground set $\bigcup_jU_j$ and independent sets
\[
 \left\{S\subseteq\bigcup_jU_j:
       S\cap U_j\in\Ical(N_j)\text{ for every }j\right\},
\]
where $\Ical(N)$ denotes the independent sets of a matroid $N$.

\subsection{Values and arrival order}

Each element $i\in E$ has a nonnegative, almost surely finite random
value $X_i$ with known law $D_i$. The variables $(X_i)_{i\in E}$ are
mutually independent. Write $D=(D_i)_{i\in E}$ and, for $S\subseteq E$,
\[
 X(S)=\sum_{i\in S}X_i,
 \qquad \OPT_M(X)=\max_{I\in\Ical}X(I).
\]
Assume $\E[\OPT_M(X)]<\infty$.

Every element arrives exactly once. At its arrival, the algorithm
observes its identity and value and irrevocably accepts or rejects it.
The accepted set must be independent after every arrival. Before the
first arrival, an almighty ordering adversary observes the entire
realization $X$ and chooses a permutation $\pi$ of $E$. Let
$\mathfrak S_E$ be the set of these permutations, and let $\ALG(X,\pi)$
be the sum of the values accepted by a specified deterministic algorithm.
Its worst-order expected reward is
\[
 \E_X\!\left[\min_{\pi\in\mathfrak S_E}\ALG(X,\pi)\right].
\]

Delete all loops before preprocessing and reject them if they arrive.
This does not change $\OPT_M(X)$. Each remaining singleton is independent,
so $\E X_i\leq\E[\OPT_M(X)]<\infty$ for every remaining element.
Hereafter $M$ and $E$ denote this loopless restriction and its ground set.
If $E=\varnothing$, use the empty set as the output. 

\subsection{The ex-ante relaxation}

For a vector $z\in\R^E$, write $z(A)=\sum_{i\in A}z_i$. Define
\[
 P(M)=\{z\in\R_{\geq0}^E:
             z(A)\leq r_M(A)\text{ for every }A\subseteq E\}.
\]
The ex-ante benchmark is
\begin{equation}
 \exante(M,D)=
 \sup_{\substack{a_i:\R_{\geq0}\to[0,1]\text{ measurable}\ (i\in E)\\
                  (\E a_i(X_i))_{i\in E}\in P(M)}}
       \sum_{i\in E}\E[X_i a_i(X_i)].
 \label{eq:exante-definition}
\end{equation}
Allowing $a_i$ to take values in $[0,1]$ includes fractional selection
at atoms of $D_i$. Since the remaining values are integrable,
$\exante(M,D)\leq\sum_i\E X_i<\infty$.
Deleting loops leaves this benchmark unchanged: the constraint for a
loop singleton forces $\E a_i(X_i)=0$, and hence $a_i(X_i)=0$ almost
surely.

\begin{lemma}
\label{lem:exante-offline}
For any nonnegative random value vector integrable on nonloops,
\[
 \exante(M,D)\geq\E[\OPT_M(X)].
\]
This inequality does not require independence of the values.
\end{lemma}

\begin{proof}
Fix an ordering of the finite family $\Ical$ and let $I^*(X)$ be the
first set attaining $\OPT_M(X)$. For each $i$, choose a measurable
function $a_i:\R_{\geq0}\to[0,1]$ such that
\[
 a_i(X_i)=\E[\ind_{\{i\in I^*(X)\}}\mid X_i]
 \quad\text{almost surely}.
\]
For every $A\subseteq E$,
\[
 \sum_{i\in A}\E a_i(X_i)
 =\E[|I^*(X)\cap A|]\leq r_M(A).
\]
These functions are feasible in~\eqref{eq:exante-definition}. Conditional
expectation gives
\[
 \sum_i\E[X_i a_i(X_i)]
 =\sum_i\E[X_i\ind_{\{i\in I^*(X)\}}]
 =\E[\OPT_M(X)].
\]
\end{proof}

For $u\in\R$, put $(u)_+=\max\{u,0\}$. Define
\begin{equation}
 e_i(t)=\E[(X_i-t)_+],\qquad
 e(A,t)=\sum_{i\in A}e_i(t)
 \quad(t\geq0,\ A\subseteq E).
 \label{eq:excess}
\end{equation}
These functions always use the original marginal laws $D_i$.
Each $e_i$ is finite, nonincreasing, and 1-Lipschitz, since
$|(x-s)_+-(x-t)_+|\leq|s-t|$.

\section{Decomposition and selection rule}
\label{sec:decomposition}

\subsection{Construction of the blocks}

Let $N$ be a loopless matroid on a nonempty subset $W\subseteq E$.
For every nonempty $A\subseteq W$, define $\tau_N(A)$ by
\begin{equation}
 e(A,\tau_N(A))=\tau_N(A)r_N(A),\qquad \tau_N(A)\geq0.
 \label{eq:root}
\end{equation}
This root exists and is unique. Indeed, $r_N(A)>0$, and
$e(A,t)-tr_N(A)$ is continuous and strictly decreasing in $t$,
is nonnegative at zero, and tends to $-\infty$ as $t\to\infty$.
For a specified $t$, call $A$ tight in $N$ if $e(A,t)=tr_N(A)$.

\begin{algorithm}[htbp]
\caption{Decomposition into matroid minors}
\label{alg:decomposition}
\textbf{Input:} The loopless matroid $M$ and the marginal laws $D_i$.
\begin{enumerate}[leftmargin=2em,label=\arabic*.]
\item Set $F_0=\varnothing$, $N=M$, and $j=1$.
\item While the ground set $W$ of $N$ is nonempty:
  \begin{enumerate}[leftmargin=2em,label=(\alph*)]
  \item Set $t_j=\max_{\varnothing\ne A\subseteq W}\tau_N(A)$ and
  \[
    U_j=\bigcup\{A\subseteq W:e(A,t_j)=t_jr_N(A)\}.
  \]
  \item Set $N_j=N|U_j$, $d_j=r_N(U_j)$, and $F_j=F_{j-1}\cup U_j$.
  \item Replace $N$ by $N/U_j$ and increase $j$ by one.
  \end{enumerate}
\item Let $m$ be the number of iterations. Return
$(U_j,F_j,N_j,d_j,t_j)_{j=1}^m$ and $\widehat M=\bigoplus_{j=1}^mN_j$.
Assign each item $i\in U_j$ the threshold $t_i=t_j$.
\end{enumerate}
\end{algorithm}

\begin{lemma}
\label{lem:decomposition}
Algorithm~\ref{alg:decomposition} is well defined and partitions $E$
into $m\leq r_M(E)$ nonempty blocks. Its prefixes $F_j$ are flats of
$M$, and
\[
 N_j=(M/F_{j-1})|U_j,
 \qquad d_j=r_M(F_j)-r_M(F_{j-1})>0.
\]
For each block,
\begin{align}
 e(A,t_j)&\leq t_jr_{N_j}(A) &&(A\subseteq U_j),
 \label{eq:block-domination}\\
 e(U_j,t_j)&=t_jd_j.
 \label{eq:block-tightness}
\end{align}
The thresholds satisfy
\begin{equation}
 t_1\geq t_2\geq\cdots\geq t_m\geq0.
 \label{eq:threshold-order}
\end{equation}
\end{lemma}

\begin{proof}
Consider an iteration with loopless residual matroid $N$ on $W$ and
$t=\max_{A\ne\varnothing}\tau_N(A)$. The definition of the roots implies
\[
 f(A):=tr_N(A)-e(A,t)\geq0\qquad(A\subseteq W).
\]
The set function $e(\cdot,t)$ is modular, meaning that
$e(A,t)+e(B,t)=e(A\cup B,t)+e(A\cap B,t)$.
Hence $f$ is submodular. If $f(A)=f(B)=0$, then
\[
 0\leq f(A\cup B)+f(A\cap B)\leq f(A)+f(B)=0.
\]
Thus tight sets are closed under union and intersection. A nonempty
tight set exists because the maximum defining $t$ is attained.
Their union $U$ is therefore nonempty and tight, and is the unique
largest tight set.

If $A$ is tight and $C=\cl_N(A)$, nonnegativity of the excesses gives
\[
 tr_N(A)=e(A,t)\leq e(C,t)\leq tr_N(C)=tr_N(A).
\]
Thus $C$ is tight. Applying this to $U$ proves $\cl_N(U)=U$.
Consequently $N/U$ is loopless. Starting with the flat
$F_0=\varnothing$ in the loopless matroid $M$, the contraction
closure formula shows inductively that each $F_j$ is a flat of $M$.
Successive contractions give $N=M/F_{j-1}$ at iteration $j$, so the
displayed formulas for $N_j$ and $d_j$ follow. The inequalities
$f(A)\geq0$ on $U_j$ and the equality $f(U_j)=0$ prove
\eqref{eq:block-domination} and~\eqref{eq:block-tightness}.

For $B\subseteq W\setminus U$, subtract the equality for $U$ from
the inequality for $U\cup B$ to obtain
\[
 e(B,t)\leq t\bigl(r_N(U\cup B)-r_N(U)\bigr)=tr_{N/U}(B).
\]
For nonempty $B$, strict monotonicity of its root-defining function
implies $\tau_{N/U}(B)\leq t$. This proves~\eqref{eq:threshold-order}.
If $t=0$, the singleton inequalities imply $\E X_i=0$ for every
$i\in W$, so every set is tight and $U=W$.
In every iteration, $U$ is nonempty in a loopless matroid and hence
has positive rank. Each contraction reduces the residual rank by at
least one. The process therefore terminates in at most $r_M(E)$
iterations. The extracted sets are disjoint and exhaust $E$.
\end{proof}

\subsection{Online selection}

\begin{algorithm}[htbp]
\caption{Online selection}
\label{alg:online}
\textbf{Preprocessing:} Run Algorithm~\ref{alg:decomposition} and set
$S_j=\varnothing$ for every block $j$.
\begin{enumerate}[leftmargin=2em,label=\arabic*.]
\item Reject every original loop.
\item When a nonloop item $i\in U_j$ arrives with value $x_i$, accept it
if and only if $x_i\geq t_j$ and $S_j\cup\{i\}\in\Ical(N_j)$.
On acceptance, replace $S_j$ by $S_j\cup\{i\}$.
\item Return $S=\bigcup_{j=1}^mS_j$.
\end{enumerate}
\end{algorithm}

For a realization $X$, define
\[
 R_j(X)=\{i\in U_j:X_i\geq t_j\},
 \qquad
 \Acal(X)=\{S\in\Ical(\widehat M):X_i\geq t_i\text{ for every }i\in S\}.
\]
Write $\Acal_{\max}(X)$ for the inclusion-wise maximal members of
$\Acal(X)$. The empty set belongs to $\Acal(X)$, so this finite family
always has a maximal member.

\begin{lemma}
\label{lem:feasibility}
Every independent set of $\widehat M$ is independent in $M$.
A set $S$ belongs to $\Acal_{\max}(X)$ if and only if
$S\cap U_j\in\Bcal_{N_j}(R_j(X))$ for every $j$.
For every realization and every arrival order, Algorithm~\ref{alg:online}
is feasible after every arrival and returns a member of
$\Acal_{\max}(X)$.
\end{lemma}

\begin{proof}
Let $A\subseteq F_{j-1}$ be independent in $M$ and let
$B\subseteq U_j$ be independent in $M/F_{j-1}$. Diminishing rank
increments and the contraction formula give
\[
 r_M(A\cup B)-r_M(A)
 \geq r_M(F_{j-1}\cup B)-r_M(F_{j-1})=|B|.
\]
Since $r_M(A)=|A|$ and $A\cap B=\varnothing$, the set $A\cup B$ is
independent. Induction over $j$ proves that every independent set of
$\widehat M$ is independent in $M$.

Membership in $\Acal(X)$ is equivalent to requiring that, for each
$j$, $S\cap U_j$ be an independent subset of $R_j(X)$. Such a set is
maximal precisely when each intersection is maximal in its block,
which is the stated basis condition.

The online rule maintains independence in each block. If an eligible
item is rejected, it lies in the closure of the accepted set in its
block at that time. Monotonicity of closure implies that it remains
in the closure of the final accepted set. Thus the final set in each
block is a basis of $R_j(X)$. The first assertion gives feasibility
after every arrival, and the second gives the claimed maximality.
\end{proof}

\section{Expected reward}
\label{sec:basis}

\subsection{A bound for the minimum-weight eligible basis}

\begin{lemma}
\label{lem:forced}
Let $N$ be a finite matroid on $W$, let $(X_i)_{i\in W}$ be independent
nonnegative integrable random variables, and fix $t\geq0$. Put
$w_i=\E[(X_i-t)_+]$, $w(A)=\sum_{i\in A}w_i$, and
$R=\{i\in W:X_i\geq t\}$. If
\begin{equation}
 w(A)\leq tr_N(A)\qquad(A\subseteq W),
 \label{eq:local-density}
\end{equation}
then
\begin{equation}
 \E\!\left[\min_{B\in\Bcal_N(R)}X(B)\right]\geq w(W).
 \label{eq:local-guarantee}
\end{equation}
More generally, for deterministic weights $a_i$ satisfying
\begin{equation}
 0\leq a_i\leq\E[(X_i-t)_+],\qquad
 a(A):=\sum_{i\in A}a_i\leq tr_N(A)\quad(A\subseteq W),
 \label{eq:dominated-weights}
\end{equation}
the same expected minimum is at least $a(W)$.
\end{lemma}

\begin{proof}
Let $F=\cl_N(R)$ and, for each $i\in W$, define the event
\[
 H_i=\{i\notin\cl_N(R\setminus\{i\})\}.
\]
The set $R\setminus\{i\}$ depends only on $(X_k)_{k\ne i}$.
Thus $H_i$ is independent of $X_i$.
On the event $\{X_i>t\}\cap H_i$, every basis of $R$ contains $i$:
otherwise a basis $B\subseteq R\setminus\{i\}$ would satisfy
$i\in R\subseteq\cl_N(B)\subseteq\cl_N(R\setminus\{i\})$,
contrary to $H_i$.

Every basis $B$ of $R$ has $r_N(F)$ elements, all with value at least
$t$. Therefore, for every realization, simultaneously for all these bases,
\begin{equation}
 \min_{B\in\Bcal_N(R)}X(B)
 \geq tr_N(F)+\sum_{i\in W}(X_i-t)_+\ind_{H_i}.
 \label{eq:forced-pointwise}
\end{equation}
Here $\ind_H$ is the indicator of the event $H$.
Taking expectations and using independence gives
\[
 \E\!\left[\min_{B\in\Bcal_N(R)}X(B)\right]
 \geq t\E[r_N(F)]+\sum_{i\in W}w_i\Prb(H_i).
\]
We prove the general assertion. Since $a_i\leq w_i$ and
$\cl_N(R\setminus\{i\})\subseteq F$,
\[
 \sum_iw_i\Prb(H_i)
 \geq\sum_i a_i\Prb(H_i)
 \geq\sum_i a_i\Prb(i\notin F)
 =a(W)-\E[a(F)].
\]
The subset inequalities in~\eqref{eq:dominated-weights} hold for every
realization of $F$, so $a(F)\leq tr_N(F)$. Substitution proves the
bound $a(W)$. Taking $a_i=w_i$ proves~\eqref{eq:local-guarantee}.
The argument also covers $t=0$, values equal to $t$, and $R=\varnothing$.
\end{proof}

\subsection{Comparison with the ex-ante benchmark}

For the decomposition in Section~\ref{sec:decomposition}, define
\begin{equation}
 K=\sum_{j=1}^m t_jd_j=\sum_{i\in E}e_i(t_i),
 \label{eq:budget}
\end{equation}
where the equality follows from~\eqref{eq:block-tightness}.

\begin{lemma}
\label{lem:budget}
For every $z\in P(M)$, $\sum_{i\in E}t_i z_i\leq K$.
\end{lemma}

\begin{proof}
The assertion is immediate if $E=\varnothing$. Otherwise put
$t_{m+1}=0$. The nested prefixes $F_j=\bigcup_{k\leq j}U_k$ and the
nonincreasing thresholds give
\begin{align*}
 \sum_{i\in E}t_i z_i
 &=\sum_{j=1}^m(t_j-t_{j+1})z(F_j)\\
 &\leq\sum_{j=1}^m(t_j-t_{j+1})r_M(F_j)\\
 &=\sum_{j=1}^m t_j\bigl(r_M(F_j)-r_M(F_{j-1})\bigr)=K.
\end{align*}
The inequality uses $z(F_j)\leq r_M(F_j)$ and
$t_j-t_{j+1}\geq0$ for each $j$.
\end{proof}

\begin{theorem}
\label{thm:master}
For the independent values and matroid specified in
Section~\ref{sec:preliminaries},
\begin{equation}
 \E\!\left[\min_{S\in\Acal_{\max}(X)}X(S)\right]
 \geq K\geq\frac12\exante(M,D).
 \label{eq:master}
\end{equation}
\end{theorem}

\begin{proof}
By Lemma~\ref{lem:feasibility}, a maximal eligible set is a union of
one basis of $R_j(X)$ in each block. Every choice of these bases is
permitted. Hence, pointwise,
\[
 \min_{S\in\Acal_{\max}(X)}X(S)
 =\sum_{j=1}^m\min_{B\in\Bcal_{N_j}(R_j(X))}X(B).
\]
Apply Lemma~\ref{lem:forced} to each $N_j$ at threshold $t_j$.
Its hypothesis is~\eqref{eq:block-domination}, so
\[
 \E\!\left[\min_{S\in\Acal_{\max}(X)}X(S)\right]
 \geq\sum_{j=1}^m e(U_j,t_j)=K.
\]

To prove the second inequality, take any feasible family $(a_i)$
in~\eqref{eq:exante-definition} and set $z_i=\E a_i(X_i)$.
For every $i$, the bounds $0\leq a_i\leq1$ imply
\[
 X_i a_i(X_i)\leq t_i a_i(X_i)+(X_i-t_i)_+.
\]
Taking expectations, summing over $i$, and using
Lemma~\ref{lem:budget} yields
\[
 \sum_i\E[X_i a_i(X_i)]
 \leq\sum_i t_i z_i+\sum_i e_i(t_i)\leq 2K.
\]
Taking the supremum over the feasible families gives
$\exante(M,D)\leq2K$.
\end{proof}

\begin{theorem}[Prophet inequality against an almighty ordering adversary]
\label{thm:prophet-complete}
Let $M$ be a finite matroid, and let its element values be independent,
nonnegative, and almost surely finite, with known marginal laws $D_i$
and $\E[\OPT_M(X)]<\infty$. Algorithm~\ref{alg:online} is deterministic,
is feasible after every arrival, and satisfies
\[
 \E_X\!\left[\min_{\pi\in\mathfrak S_E}\ALG(X,\pi)\right]
 \geq\frac12\exante(M,D)
 \geq\frac12\E[\OPT_M(X)].
\]
Its preprocessing depends only on $M$ and the laws $D_i$.
\end{theorem}

\begin{proof}
Deleting loops preserves both benchmarks, and the algorithm rejects
them in every order. It therefore suffices to consider the loopless
restriction. Algorithm~\ref{alg:decomposition} is deterministic because
each threshold and largest tight set is uniquely specified. All its
inputs are the matroid and the marginal laws. The online decisions use
only these precomputed quantities, the previously accepted sets, and
the arriving item's identity and value.

Lemma~\ref{lem:feasibility} proves feasibility and shows that, for every
realization and permutation, the terminal set belongs to
$\Acal_{\max}(X)$. Consequently,
\[
 \min_{\pi\in\mathfrak S_E}\ALG(X,\pi)
 \geq\min_{S\in\Acal_{\max}(X)}X(S).
\]
Taking expectations and applying Theorem~\ref{thm:master} and
Lemma~\ref{lem:exante-offline} proves the result. All minima are over
finite families with measurable eligibility and selection conditions.
Their values are measurable and bounded by the integrable offline
optimum; thus all displayed expectations are well defined.
\end{proof}

\section{Correlated delegation}
\label{sec:delegation}

We now turn the eligible-set construction into a delegation mechanism,
proving the one-half ex-ante guarantee of Theorem~\ref{thm:delegation}.
The principal commits to a menu before utilities are realized; the
agent then chooses a proposal from this menu after observing all
utilities.

The menu uses the same blocks and thresholds as the online algorithm.
A proposal is accepted precisely when every submitted item meets its
threshold and the submitted set is independent in $\widehat M$.
These conditions depend only on submitted identities and principal
values, so the principal can verify them without observing agent
utilities or the values of omitted items. The resulting mechanism
is specified in Algorithm~\ref{alg:delegation}.

\begin{algorithm}[htbp]
\caption{Proposal-only delegation mechanism}
\label{alg:delegation}
\textbf{Preprocessing:} Run Algorithm~\ref{alg:decomposition} using the
principal-value laws and announce the resulting blocks, thresholds,
and residual matroid constraints.
\begin{enumerate}[leftmargin=2em,label=\arabic*.,itemsep=2pt]
\item Given a proposal $S$ with verifiable principal values,
      reject it if it contains an original loop or any
      item $i$ with $X_i<t_i$.
\item Reject if $S\cap U_j\notin\Ical(N_j)$ for any block $j$.
\item Otherwise, implement every item in $S$.
      Rejection implements the empty set. The empty proposal is allowed.
\end{enumerate}
\end{algorithm}

Theorem~\ref{thm:master} gives the required expected-value bound even
when the least valuable maximal eligible set is chosen in each
realization. To apply it to this menu, we show that
strictly positive agent utilities make every best-response outcome
maximal, regardless of how the agent breaks ties.

\begin{proof}[Proof of Theorem~\ref{thm:delegation}]
The proposals accepted by Algorithm~\ref{alg:delegation} in realization
$X$ are exactly the sets in $\Acal(X)$, each of which is independent
in $M$ by Lemma~\ref{lem:feasibility}. The mechanism uses only $M$
and the marginal principal-value laws, so it is deterministic and
agent-oblivious.

Fix a realization $(X,Y)$ and let $S$ be the implemented outcome of
any best response. Then $S\in\Acal(X)$: an accepted proposal belongs
to this family, and rejection or nonparticipation implements the
empty set, which also belongs to it. If $S$ were not maximal in
$\Acal(X)$, some eligible item $i\notin S$ could be added while
preserving independence in $\widehat M$. The proposal $S\cup\{i\}$
would therefore be accepted and give the agent utility
$Y(S)+Y_i>Y(S)$, contradicting optimality. Thus every best-response
outcome belongs to $\Acal_{\max}(X)$, and hence
\[
 \min_{S\in\BR(X,Y)}X(S)
 \geq\min_{S\in\Acal_{\max}(X)}X(S).
\]
Taking expectations and applying Theorem~\ref{thm:master} proves
\eqref{eq:main-delegation}. The maximality argument holds for each
realization and uses only $Y_i>0$, so arbitrary dependence of agent
utilities on principal values and on one another is allowed.
\end{proof}

\paragraph{Why strictly positive agent utilities are assumed.}
With one deterministic item $(X,Y)=(1,0)$, the empty proposal is a
best response to every menu. Under ties resolved against the principal,
no positive approximation guarantee is possible. Thus strict positivity
in Theorem~\ref{thm:delegation} is substantive.

\section{Polynomial-time implementation}
\label{sec:implementation}

We implement the decomposition exactly for explicitly listed rational
finite-support distributions, using an independence oracle for $M$.
Let $n$ be the number of nonloop items and $L\geq1$ the total binary
length of the support lists and item labels, taking $L=1$ for an empty
input.

The implementation has three tasks: simulate the residual matroids,
find each maximum surplus threshold, and recover its largest tight
flat. Sections~\ref{sec:implementation-minors}--\ref{sec:implementation-flat}
address these tasks in order. Section~\ref{sec:implementation-complexity}
assembles them and establishes the bit complexity and online query
bound. Section~\ref{sec:implementation-delegation} then describes
proposal verification and best-response computation.

\begin{theorem}[Running time]
\label{thm:implementation}
For rational finite-support distributions, Algorithm~\ref{alg:decomposition}
can be implemented exactly in polynomial bit time with polynomially many
independence queries and $O(n(L+\log\max\{1,n\}))$ submodular
minimizations. Thereafter, Algorithm~\ref{alg:online} uses at most
one original independence query per eligible arrival and none otherwise.
\end{theorem}

Polynomial time here includes exact rational arithmetic and writing
oracle queries; it is not merely a bound on the number of real-arithmetic
operations. If an original independence query takes time $T_M$,
the preprocessing time is polynomial in $n$, $L$, and $T_M$. We do not
claim a strongly polynomial bound or linear total running time for the
online phase. For general integrable distributions,
Theorem~\ref{thm:prophet} is an existence result unless a representation
supporting the required expectation computations is supplied.

\paragraph{Input representation.}
Items have distinct nonempty binary labels. For each item $i$, the
input lists rational values $v_{ia}\geq0$ and probabilities
$p_{ia}\geq0$ such that $\Pr[X_i=v_{ia}]=p_{ia}$ and
$\sum_a p_{ia}=1$. Rationals are written in lowest terms
with positive denominators; integer bit lengths count zero as one bit.
Discard zero-probability entries and combine repeated support values
within each item. These normalizations take polynomial time and preserve
the distributions. In all definitions below, the support lists refer to
these normalized lists. Remove loops with one independence query per
original item. If $n=0$, the algorithm returns the empty set.
Below, assume $n\geq1$.

\subsection{Implementing the residual matroids}
\label{sec:implementation-minors}

Store a basis $C_j$ of every prefix flat $F_j$, with $C_0=\varnothing$.
If $C$ is a basis of $F$ and $A\subseteq E\setminus F$, then
\begin{equation}
 A\in\Ical(M/F)\quad\Longleftrightarrow\quad C\cup A\in\Ical(M).
 \label{eq:implementation-contraction}
\end{equation}
Indeed, $C$ and $F$ have the same closure, and hence
\[
 r_{M/F}(A)=r_M(F\cup A)-r_M(F)
           =r_M(C\cup A)-|C|.
\]
Thus a residual independence query costs one original query. To compute
$r_{M/F}(A)$, start with $C$ and greedily insert elements of $A$ whenever
independence is preserved. The number of insertions is the desired rank,
and the scan uses at most $|A|$ original queries. Restrictions are handled
by restricting the allowed query elements.

After extracting $U_j$, greedily extend $C_{j-1}$ using the elements of
$U_j$ to obtain $C_j$. These extensions use at most $n$ original queries
over the entire decomposition. The stored bases are auxiliary sets for
simulating contractions. Their elements need not be selected online,
and their realized values are never used in preprocessing.

\subsection{Computing an exact threshold}
\label{sec:implementation-threshold}

We first use submodular minimization to decide whether a candidate
value lies below the maximum threshold. Rational separation then
lets us recover an exact maximizing set from a sufficiently short
bisection interval; its threshold is obtained by solving a single
piecewise-linear equation.

\paragraph{A decision oracle.}
Fix a nonempty residual ground set $W$ and the loopless residual matroid
$N=M/F$. For a rational $\lambda\geq0$, define
\[
 e_i(\lambda)=\sum_a p_{ia}(v_{ia}-\lambda)_+,
 \qquad
 f_\lambda(A)=\lambda r_N(A)-\sum_{i\in A}e_i(\lambda).
\]
The function $f_\lambda$ is submodular. Its value oracle first computes
the excesses from the support lists and then answers each set query
using one residual rank computation and a sum. In particular, a value
query uses at most $n$ original independence queries. We write
$\SFM(f)$ for a polynomial-time submodular minimization routine that
returns $(f(A),A)$ with $A\subseteq W$ and
$f(A)=\min_{B\subseteq W}f(B)$ for a rational-valued submodular
function on $2^W$~\cite{schrijver2000}.

Let
$\tau^*=\max_{\varnothing\ne A\subseteq W}\tau_N(A)$, where
$e(A,\tau_N(A))=\tau_N(A)r_N(A)$. Strict monotonicity of the
root-defining function gives the decision rule
\begin{equation}
 \tau^*>\lambda
 \quad\Longleftrightarrow\quad
 \min_{A\subseteq W}f_\lambda(A)<0.
 \label{eq:implementation-decision}
\end{equation}
The empty set has value zero, so a negative minimum always supplies a
nonempty witness. A zero minimum means $\tau^*\leq\lambda$, including
the case of equality.

\paragraph{Separation of candidate thresholds.}
Let $D$ be the product, with multiplicity, of the reduced denominators
of all probabilities and values in the normalized input lists, and
put $Q=2nD$. Then $\log D=O(L)$. Let $V$ be the largest support
value in the normalized lists of all original items. The following
separation bound converts the decision procedure into an exact algorithm.

\begin{lemma}[Separation of candidate thresholds]
\label{lem:implementation-precision}
Every candidate threshold $\tau_N(A)$ in a loopless minor of $M$
is rational with reduced denominator at most $Q$. Any two distinct
candidate thresholds differ by at least $Q^{-2}$.
\end{lemma}

\begin{proof}
At a root $\tau$, let
\[
 \alpha=\sum_{i\in A}\sum_{a:v_{ia}>\tau}p_{ia}v_{ia},
 \qquad
 \beta=\sum_{i\in A}\sum_{a:v_{ia}>\tau}p_{ia}.
\]
Atoms at $\tau$ contribute zero. Since $N$ is loopless and $A$ is
nonempty, $r_N(A)>0$, and the root equation yields
\[
 \tau=\frac{D\alpha}{D r_N(A)+D\beta}.
\]
Both numerator and denominator are integers. The denominator is
positive and at most $2nD$, proving the first claim and the separation
bound.
\end{proof}

\paragraph{Recovering an exact maximizing set.}
If all residual means vanish, the maximum threshold is zero and the
entire residual ground set is tight. Otherwise $\tau^*>0$. With the global $V$ just defined, start with $\ell=0$ and $u=V$.
Bisect at $\lambda=(\ell+u)/2$, setting $\ell=\lambda$ if the
minimum in~\eqref{eq:implementation-decision} is negative and
$u=\lambda$ otherwise. After every step this maintains
\[
 \ell<\tau^*\leq u.
\]
After $O(L+\log n)$ bisections, $u-\ell<1/Q^2$. Every minimizing set
$A$ for $f_\ell$ is nonempty, has negative value, and satisfies
\[
 \ell<\tau_N(A)\leq\tau^*\leq u.
\]
Lemma~\ref{lem:implementation-precision} forces
$\tau_N(A)=\tau^*$. Thus the procedure recovers an exact maximizing
witness, rather than merely an approximate threshold.

\paragraph{Solving for the witness's threshold.}
To compute the witness's root, sort the distinct values among $0$, $V$,
and the support values of items in $A$ into
$0=s_0<s_1<\cdots<s_k=V$. If $V=0$, return zero.
On each closed interval $[s_r,s_{r+1}]$, set
\[
 \alpha=\sum_{i\in A}\sum_{a:v_{ia}>s_r}p_{ia}v_{ia},
 \qquad
 \beta=\sum_{i\in A}\sum_{a:v_{ia}>s_r}p_{ia}.
\]
Then $e(A,t)=\alpha-\beta t$ throughout that interval, including
its endpoints. The candidate root is $\alpha/(r_N(A)+\beta)$.
Scan the intervals in increasing order, updating the tail sums, and
return the first candidate lying in its closed interval and satisfying
$e(A,t)=tr_N(A)$. Such a candidate always exists: the unique root lies
in $[0,V]$, and the affine formulas agree at shared endpoints.
There are at most as many intervals as support entries plus one,
so the subroutine takes polynomial time. Denote it by
$\operatorname{ExactRoot}(N,A)$.

\subsection{Recovering the largest tight flat}
\label{sec:implementation-flat}

At $t=\tau^*$, the function $f_t$ is nonnegative and its zero sets are
closed under union and intersection. Their union is therefore the
unique largest tight set $U$. It is a flat: closing a tight set does
not increase its rank, while the nonnegative excess weights cannot
decrease; nonnegativity of $f_t$ then forces its closure to remain tight.

There is a direct way to select $U$ with one further submodular
minimization. Write $t=a/b$ in lowest terms and let $q=|W|$. Minimize
\begin{equation}
 g(A)=(q+1)Db f_t(A)-|A|.
 \label{eq:implementation-canonical}
\end{equation}
The function $g$ is submodular. Since $Db f_t(A)$ is a nonnegative
integer, every nontight set has $g(A)\geq q+1-|A|\geq1$.
Every tight set has $g(A)=-|A|\leq0$. A minimizer of $g$ is therefore
a tight set of maximum cardinality, which must be $U$. This implements
the prescribed tie rule exactly, without a numerical perturbation.

\subsection{Complete algorithm and running time}
\label{sec:implementation-complexity}

Algorithm~\ref{alg:exact-block} combines threshold recovery with the
selection of the largest tight flat. Applying it successively to the
residual matroids implements Algorithm~\ref{alg:decomposition}.

\begin{algorithm}[htbp]
\caption{Exact extraction of one block}
\label{alg:exact-block}
\textbf{Input:} Loopless residual matroid $N$ on
$W\ne\varnothing$; normalized support lists; the routine $\SFM$
specified above; the denominator product $D$ defined above,
$Q=2nD$, and the largest normalized support value $V$ across all original items.\par
\textbf{Output:} Maximum surplus threshold $t$ and its largest tight flat $U$.
\begin{enumerate}[leftmargin=2em,label=\arabic*.,itemsep=2pt]
\item \textbf{If} $\E X_i=0$ for every $i\in W$, \textbf{return} $(0,W)$.
\item Set $\ell\gets0$ and $u\gets V$.
\item \textbf{While} $Q^2(u-\ell)\geq1$:
  \begin{enumerate}[leftmargin=2em,label=(\alph*),itemsep=2pt]
  \item Set $\lambda\gets(\ell+u)/2$ and compute
        $(m,A)\gets\SFM(f_\lambda)$.
  \item \textbf{If} $m<0$, set $\ell\gets\lambda$;
        \textbf{otherwise}, set $u\gets\lambda$.
  \end{enumerate}
\item Compute $(m,A)\gets\SFM(f_\ell)$.
      The minimum is negative and $A$ maximizes the root.
\item Set $t\gets\operatorname{ExactRoot}(N,A)$ and write $t=a/b$
      in lowest terms with $b\geq1$.
\item Define $g(B)=(|W|+1)Db f_t(B)-|B|$ for $B\subseteq W$;
      compute $(m_g,U)\gets\SFM(g)$.
\item \textbf{Return} $(t,U)$.
\end{enumerate}
\end{algorithm}

\paragraph{Bit complexity of the value oracles.}
The bisection in Section~\ref{sec:implementation-threshold} halves
an interval of initial length $V$ until its length is less than
$1/Q^2$. It uses at most
$\lceil\log_2(\max\{1,V\}Q^2)\rceil+1=O(L+\log n)$ iterations,
and every endpoint is a dyadic multiple of $V$. Its numerator and
denominator therefore have $O(L+\log n)$
digits. If $b_\lambda$ is the denominator of a query $\lambda$, then
$Db_\lambda f_\lambda(A)$ is an integer. For
$0\leq\lambda\leq V$, we have
$|f_\lambda(A)|\leq2nV$, so this integer has $O(L+\log n)$ digits.
The same bound holds at the recovered root, whose denominator is at
most $Q$, and after the additional scaling used to recover the largest
tight set: both $Db_\lambda f_\lambda(A)$ and
$(|W|+1)Db_\lambda f_\lambda(A)-|A|$ are integers with
$O(L+\log n)$ digits. Thus the exact value oracles have
polynomial-size outputs.

\begin{proof}[Proof of Theorem~\ref{thm:implementation}]
Run Algorithm~\ref{alg:decomposition}, using
Algorithm~\ref{alg:exact-block} for each extraction and maintaining
the stored prefix bases. Every stage contracts a nonempty flat of
positive rank, so there are at most $r_M(E)\leq n$ stages. A nonzero
stage uses $O(L+\log n)$ bisection minimizations, one witness
minimization, and one minimization for the canonical flat. Zero
stages require no minimization. This proves the stated number of
calls. Each value oracle uses polynomial work and at most $n$
original independence queries. The precision lemma and exact
submodular minimization imply polynomial total preprocessing time.

Store an item-to-block table and $C_{j-1}$ for every block. When an
eligible item $i\in U_j$ arrives, query the original oracle on
\[
 C_{j-1}\cup S_j\cup\{i\}.
\]
By~\eqref{eq:implementation-contraction}, this is exactly the
independence test required by Algorithm~\ref{alg:online}. The query
set can be constructed in $O(n)$ set operations. Each nonloop
arrival requires one exact threshold comparison, whose bit cost
depends on the encoding length of the arriving value. For draws from
the input supports this length is bounded by $L$. The implementation
agrees with Algorithm~\ref{alg:online} for every nonnegative rational
realization and every arrival order. It makes no independence query
on loops or ineligible items and at most one per eligible item,
hence at most $n$ in total. If $n=0$, it returns the empty set.
Since the implementation produces precisely the thresholds,
blocks, and online decisions of the mathematical construction,
Theorem~\ref{thm:prophet} applies without any loss in the guarantee.
\end{proof}

\subsection{Delegation verification and best responses}
\label{sec:implementation-delegation}

\paragraph{Verifying a proposal.}
The same stored bases implement the delegation verifier. First reject
any proposal containing an original loop. Check each submitted value
against its threshold, and for each nonempty submitted block query $C_{j-1}\cup(S\cap U_j)$ in the original matroid. This uses
at most one original independence query per nonempty submitted block.
The verifier inspects neither omitted realized values nor agent utilities.

\paragraph{Computing a best response.}
For fixed nonnegative realized agent utilities, delete original loops
and items with $X_i<t_i$, sort the remaining items in nonincreasing
order of $Y_i$, and apply weighted matroid greedy in $\widehat M$.
This maximizes $Y(S)$ over $\Acal(X)$, so it computes a best response
to Algorithm~\ref{alg:delegation}. Any ordering of equal-weight items
works. Using the stored prefix bases, the computation takes
$O(n\log\max\{2,n\})$ comparisons of agent utilities and at most
$n$ original independence queries. This computational statement also
allows zero agent utilities; the approximation guarantee in
Theorem~\ref{thm:delegation} requires strict positivity and holds for
every best response, regardless of how it is computed.

\paragraph{AI Use Statement.}
This work resulted from extensive interaction between the authors and
OpenAI Astra, with the authors providing suggestions and critically
checking Astra's responses through iterative exchanges. OpenAI Astra
was also used in writing and revising the manuscript through detailed
prompts from the authors, with Gemini Argon assisting in some parts.
The authors directed the process and evaluated the proposed ideas
and arguments. The results of this paper have also been formalized in
the Lean proof assistant.

\begingroup
\small
\interlinepenalty=10000
\bibliographystyle{abbrvnat}
\bibliography{delegation}
\endgroup

\end{document}